\documentclass[
    ,final            
  ]
  {aipproc}

\layoutstyle{8x11single}

\begin{document}

\newcommand{\beq}{\begin{equation}} 
\newcommand{\eeq}{\end{equation}} 
\newcommand{\beqa}{\begin{eqnarray}} 
\newcommand{\eeqa}{\end{eqnarray}} 
\renewcommand{\thefootnote}{\#\arabic{footnote}} 
\newcommand{\ve}{\varepsilon} 
\newcommand{\krig}[1]{\stackrel{\circ}{#1}}
\newcommand{\GeV}{GeV$^2$}
\newcommand{\deut}{$^2$H}
\newcommand{\pdeut}{$\stackrel{\rightarrow}{^2\rm{H}}$}
\newcommand{\he}{$^3\rm{He}$}
\newcommand{\phe}{$\stackrel{\rightarrow}{^3\rm{He}}$}
\newcommand{\pDeen}{\pdeut($\vec{e},e^\prime n$)}
\newcommand{\Deepn}{\deut($\vec{e},e^\prime \vec{n})$}
\newcommand{\Heen}{\phe($\vec{e},e^\prime n$)}
\newcommand{\Hee}{\phe($\vec{e},e^\prime$)}
\newcommand{\Ee}{\ensuremath{E_e}} 
\newcommand{\thetae}{\ensuremath{\theta_e}} 
\newcommand{\GD}{\ensuremath{G_D}}
\newcommand{\GE}{\ensuremath{G_E}}
\newcommand{\GEp}{\ensuremath{G_E^{p}}}
\newcommand{\GEpGMp}{\ensuremath{G_E^p/G_M^p}}
\newcommand{\GEn}{\ensuremath{G_E^{n}}}
\newcommand{\GM}{\ensuremath{G_M}}
\newcommand{\GMp}{\ensuremath{G_M^{p}}}
\newcommand{\GMn}{\ensuremath{G_M^{n}{}}}
\newcommand{\Q}{\ensuremath{Q^{2}{}}}
\newcommand{\updeg}{$^{o}$}
\newcommand{\vs}{\vspace{-0.0cm}} 

\def\Journal#1#2#3#4{{#1} {\bf #2}, (#4) #3}
\def\NCA{Nuovo Cimento}
\def\NIM{Nucl. Instrum. Methods}
\def\NIMA{{Nucl. Instrum. Methods} A}
\def\EPJ{{Eur. Phys. Jour.} A}
\def\JPG{J. Phys. G: Nucl. Part. Phys.}
\def\NPA{{Nucl. Phys.} A}
\def\NPB{{Nucl. Phys.} B}
\def\PLB{{Phys. Lett.}  B}
\def\PRL{Phys. Rev. Lett.}
\def\PRC{{Phys. Rev.}  C}
\def\PRD{{Phys. Rev.} D}
\def\RMP{Rev. Mod. Phys.}
\def\ZPA{{Z. Phys.} A}
\def\ZPC{{Z. Phys.} C}

\title{The Pion Cloud of the Nucleon: Facts and popular Fantasies}


\author{Ulf-G. Mei{\ss}ner}{
  address={Universit\"at Bonn, Helmholtz-Institut f\"ur Strahlen- und
  Kernphysik, D-53115 Bonn, Germany\\Forschungszentrum
  J\"ulich, Institut f\"ur Kernphysik, D-52428 J\"ulich, Germany}
}

\begin{abstract}
I discuss the concept of the pion cloud surrounding the nucleon and other
hadrons - and its limitations.
\end{abstract}

\maketitle


\section{Introduction}
\label{intro}

Long before QCD, meson theory was invented to describe the nuclear forces
\cite{Yuk}. It was quickly realized that pion-nucleon scattering data require
a large coupling constant. As a further consequence of this strong coupling, 
many virtual mesons -- Yukawas pions -- were expected to be associated with
the nucleon. This was the birth of the pion cloud, the main topic of this
contribution. Heisenberg and Wentzel developed a consistent approach to the
strong coupling limit, treating the mesons as classical fields and the finite
nucleon size providing an UV cutoff. For an early calculation of that period, 
see e.g. Ref.~\cite{FHK}. Many of these ideas have survived until
today, but now we know that low-energy QCD is governed by the spontaneous
breakdown of its chiral symmetry (for the light quarks) with the pion taking
over a special role as a (Pseudo-)Goldstone boson. In this talk, I will be
concerned with the pionic contribution to the nucleon (hadron) structure,
loosely called the ``pion cloud''. There is no doubt that is an important
part of nucleon structure, but the main questions to be addressed are:
i) is it possible to uniquely and unambiguously define the contribution 
of the ``pion cloud'' to any given observable? and ii) how could such a 
contribution be quantified? There is lots of folklore about this issue,
my goal is to be more precise and show that while this concept provides a
nice intuitive picture, it can hardly be made quantitative without resorting
to uncontrolled models. But let us go step by step. The first question can
best be addressed in the framework of chiral perturbation theory (CHPT)
(for a recent review, see~\cite{BMrev}).
To be precise, consider a single nucleon. In baryon CHPT a nucleon typically  
emits a pion, this energetically forbidden $\pi$N intermediate state lives for a 
short while and then  the pion is reabsorbed by the nucleon, in accordance
with the uncertainty principle.  This mechanism is responsible for the
venerable old idea of the ``pion cloud'' of the nucleon, which in CHPT can be put on 
the firm ground of field theoretical principles. This will be discussed in 
more detail in the next section. As will be shown, such loop contributions are
in general not scale-independent and thus can not provide the required
model-independent definition. I will then analyze the low-energy structure of the
nucleons' electromagnetic form factors and show which constraints are set
by fundamental principles (like unitarity and chiral symmetry) 
on their pionic contribution. This will then allow
for an -- albeit model-dependent -- extraction of the longest range
contribution to these fundamental nucleon structure quantities. 
Throughout this talk, I eschew models.

\section{Chiral loops as a representation of the pion cloud}
\label{sec:loops}

Beyond tree level, any observable calculated in CHPT receives contributions
from tree and loop graphs. Naively, these loop diagrams qualify as the
natural candidate for a precise definition of the ``pion cloud'' of any given
hadron. The loop graphs not only generate the imaginary
parts of the pertinent observables but are also -- 
in most cases -- divergent, requiring regularization and 
renormalization. In CHPT,  one usually chooses a mass--independent
regularization scheme to avoid power divergences (there are, however,
instances where other regulators are more appropriate or physically intuitive.
For a beautiful discussion of this and related issues,
see e.g. Refs.~\cite{Georgi:1994qn,Espriu:1993if}). 
The method of choice in CHPT
is dimensional regularization (DR), which introduces the scale $\lambda$. 
Varying this scale has no influence on any
observable $O$ (renormalization group invariance),
\beq
\frac{d}{d\lambda} \, O (\lambda) = 0~,
\eeq
but this also means that it makes little sense to assign a physical meaning to
the separate contributions from the contact terms and the loops. Physics,
however, dictates the range of scales appropriate for the process under
consideration --- describing the pion vector radius (at one loop) 
by chiral loops alone would 
necessitate a scale of about 1/2~TeV (as stressed long ago by Leutwyler). In
this case, the coupling of the $\rho$--meson generates the strength of the
corresponding one-loop counterterm that gives most of the pion radius ---
more on this below. The most intriguing aspects of chiral loops are the
so-called {\bf chiral logarithms} (chiral logs). In the chiral limit, the pion 
cloud becomes long-ranged and there is no more Yukawa factor $\sim \exp(-M_\pi
r)$ to cut it off. This generates terms  like $\log M^2_\pi, 1/ M_\pi,
\ldots$, that is contributions that are non--analytic in the quark masses.
Such statements can be applied to all hadrons that are surrounded by a cloud of pions 
which by virtue of their small masses can move away very far from the object
that generates them. Stated differently, in QCD the approach to the chiral limit
is non--analytic in the quark masses and the low--energy structure of QCD can
therefore not be analyzed in terms of a simple Taylor expansion.
The exchange of the massless Goldstone bosons generates poles and cuts 
starting at zero momentum transfer, such that the Taylor series expansion 
in powers of the momenta fails. This is a general phenomenon of theories 
that contain massless particles -- the Coulomb scattering amplitude due to 
photon exchange is proportional to $e^2/t$, with $t = (p'-p)^2$ the momentum 
transfer squared between the two charged particles. 
Let me return to the discussion of the chiral loops.
As stated before, most loops are divergent. In DR, all one--loop divergences
are simple poles in $1/(d-4)$, where $d$ is the number of space-time
dimensions.
Consequently, these divergences can be absorbed in the pertinent 
low-energy constants (LECs) that accompany the corresponding local operators
at that order in harmony with the underlying symmetries. For a given LEC $L_i$
this amounts to $L_i \to L_i^{\rm ren} + \beta_i \, {\rm L}(\lambda)~,$
where ${\rm L} \sim 1/(d-4)$ and
$\beta_i$ is the corresponding $\beta$--function. The renormalized
and finite  $L_i^{\rm ren}$ must be determined by a fit to data
(or calculated eventually using lattice QCD). Having determined the values of
the LECs from experiment, one is faced with the issue of trying to understand
these numbers. Not surprisingly, the higher mass states of QCD leave their
imprint in the LECs. Consider again the $\rho$-meson contribution to the
vector radius of the pion. Expanding the $\rho$-propagator in powers of
$t/M_\rho^2$, its first term is  a contact term of dimension four, with
the corresponding finite LEC $L_9$ given by $L_9 = F_\pi^2/2M_\rho^2 \simeq 7.2 \cdot
10^{-3}$, close to the empirical value $L_9 = 6.9  \cdot 10^{-3}$ at $\lambda
= M_\rho$. This so--called resonance saturation 
(pioneered in Refs.\cite{Ecker:1988te,Ecker:1989yg,Donoghue:1988ed})
 holds more generally for most LECs at one loop and is frequently used 
in two--loops calculations to estimate the ${\cal O}(p^6)$ LECs (for a
recent study on this issue, see~\cite{Kampf:2006bn}).
Let us now discuss the  the ``pion cloud'' of the
nucleon in the context of these considerations.
Consider as an example the isovector
Dirac radius of the proton \cite{Bernard:2003rp} (for precise definitions,
see the next section). The first loop contributions appear at third order 
in the chiral expansion, leading to
\beq\label{eq:Dirac}
\langle r^2 \rangle_1^V = \left(0.61 - \left(0.47\,{\rm GeV}^{2}\right)\,
\tilde{d} (\lambda) + 0.47 \log\frac{\lambda}{1\,{\rm GeV}} \right) ~{\rm
fm}^2~, 
\eeq
where $\tilde{d} (\lambda)$ is a dimension three pion--nucleon LEC that
parameterizes the ``nucleon core'' contribution.  
Compared to the empirical value 
$(r_1^v)^2=0.585$ fm$^2$ \cite{MMD} we note that several combinations 
of $(\lambda,\tilde d(\lambda))$  pairs can reproduce the empirical result, e.g.
\begin{eqnarray}
\left(1~\rm{GeV},+0.06~\rm{GeV}^{-2}\right),\left(0.943~\rm{GeV},
\rm{0.00~GeV}^{-2}\right),
\left(0.6~\rm{GeV},-0.46~\rm{GeV}^{-2}\right)\, .
\end{eqnarray}
An important observation to make is that even the sign of the ``core'' contribution 
to the radius can change within a reasonable range typically used for 
the scale $\lambda$.
Physical intuition would tell us that the 
value for the coupling $\tilde d$  should be negative such that the nucleon core 
gives a {\em positive} contribution to the isovector 
Dirac radius, but field theory tells us that for (quite reasonable) 
regularization scales above $\lambda=943$ MeV this need not be the case.
In essence, only the sum of the core and the cloud contribution constitutes a 
meaningful quantity that should be discussed. This observation holds for any
observable - not just for the isovector Dirac radius discussed here.

\section{Nucleon electromagnetic form factors: Basic definitions}
\label{basic}

To analyze the pion cloud contribution to the nucleons' electromagnetic
form factors in more detail, we must collect some basic definitions. 
These form factors are defined 
by the nucleon matrix element of the quark electromagnetic current, 
\beq 
\langle N(p') | \bar q \gamma^\mu {\hat Q} q | N(p)\rangle 
= \bar u (p') \left[ \gamma^\mu \, F_1 (q^2) + \frac{i}{2m} 
\sigma^{\mu\nu} (p'-p)_\nu \,  F_2 (q^2) \right] u(p)~, 
\eeq 
with $q^2 = (p'-p)^2 =t$ the invariant momentum transfer squared, 
${\hat Q}$ the quark charge matrix, and $m$ the nucleon mass. $F_1 (q^2)$ 
and $ F_2 (q^2)$ are the Dirac and the Pauli form factors, respectively. 
Following the conventions of \cite{MMD}, we decompose the form factors into 
isoscalar ($S$) and isovector ($V$) components, 
\beq 
F_i (q^2) =  F_i^S (q^2) + \tau_3 \,  F_i^V (q^2)~,\quad i=1,2\,, 
\eeq 
subject to the normalization 
$F_1^S (0) = F_1^V (0) = 1/2\, ,$ $F_2^{S,V} (0) 
= (\kappa_p \pm \kappa_n)/{2}~,$ 
with $\kappa_p \, (\kappa_n) = 1.793\, (-1.913)$ the anomalous magnetic moment 
of the proton (neutron). We will also use the Sachs form factors, 
\beq
\label{conv}
G_E^I (q^2) = F_1^I (q^2) + \frac{q^2}{4m^2} F_2^I (q^2)~, 
\quad G_M^I (q^2) = F_1^I (q^2) + F_2^I (q^2)~, \quad I = S, V ~. 
\eeq 
These are commonly referred to as the electric and the magnetic nucleon form 
factors. The slope of the form factors at $q^2=0$ can be expressed in terms 
of a nucleon radius 
\beq 
\label{radidef} 
\langle r^2\rangle_i^I = \frac{6}{F_i^I (0)}\left.
\frac{dF_i^I (q^2)}{dq^2} \right|_{q^2=0}\,,\quad i=1,2\,, \quad I = S, V\,,
\eeq 
and analogously for the Sachs form factors. The analysis of the  nucleon 
electromagnetic form factors proceeds most
directly through the spectral representation given by
\beq\label{specfct} F_i^I (q^2) = \frac{1}{\pi}\, \int_{{(\mu_0^I)}^2}^\infty 
\frac{\sigma_i^I (\mu^2) \, d\mu^2}{\mu^2 - q^2}~, \quad i =1,2~, 
\quad I = S, V~, 
\eeq 
in terms of the real {\bf spectral functions} 
$\sigma_i^I (\mu^2) = {\rm Im}\,F_i^I (\mu^2)$. 
The corresponding thresholds are given by 
$\mu_0^S = 3M_\pi$, $\mu_0^V = 2M_\pi$, with $M_\pi$ the charged pion mass.
Since the isovector spectral function is non--vanishing for smaller momentum 
transfer (starting at the two--pion cut) than the isoscalar one (starting at 
the three--pion cut), the isovector spectral functions plays a more important
role in the question of the pionic contribution to the nucleon structure. 
More precisely, let us consider the nucleon form factors in the space-like region.
In the Breit--frame (where no energy is transferred), any form factor $F$ can 
be written as the Fourier--transform of a coordinate space density, 
\beq 
F({\bf q}^2) = \int d^3{\bf r}\, {\rm e}^{i {\bf q}\cdot {\bf r}}\, 
\rho(r)~, 
\eeq 
with ${\bf q}$ the three--momentum transfer. In particular, comparison with 
Eq.~(\ref{specfct}) allows us to express the density $\rho (r)$ in terms of the 
spectral function 
\beq
\label{dens} 
\rho(r) = \frac{1}{4\pi^2}\, \int_{\mu_0^2}^\infty d\mu^2 \, 
\sigma (\mu^2) \, \frac{{\rm e}^{-\mu r}}{r}~. 
\eeq 
Note that for the electric and the magnetic Sachs form factor, $\rho (r)$ is 
nothing but the charge and the magnetization density, respectively. For
the Dirac and Pauli form factors, Eq.~(\ref{dens}) should be considered as
a formal definition. This 
equation expresses the density as a linear combination of Yukawa distributions,
each of mass $\mu$. The lightest mass hadron is the pion, and from 
Eq.~(\ref{dens}) it is evident that pions are responsible for the long--range 
part of the electromagnetic structure of the nucleon. This contribution is 
commonly called the ``pion cloud'' of the nucleon and in fact this long--range 
low--$q^2$ contribution to the nucleon form factors can be directly derived
from unitarity or be calculated  on the basis of chiral perturbation theory,
as discussed next.

\section{Spectral functions and their low-energy constraints}
\label{spec}

The spectral functions defined in Eq.~(\ref{specfct}) are the 
central quantities in the dispersion-theoretical approach. They
can in principle be constructed from experimental data. 
In practice, this program can only be carried out for 
the lightest two-particle intermediate states. Higher mass contributions
are usually parameterized in terms of vector meson poles. For the
discussion of the pion cloud, only the lightest mass (longest range) 
contributions to the spectral functions are of relevance. These will
be discussed next.

\begin{figure}[t] 
\vspace{0.9cm} 
\resizebox{0.60\textwidth}{!}{%
  \includegraphics{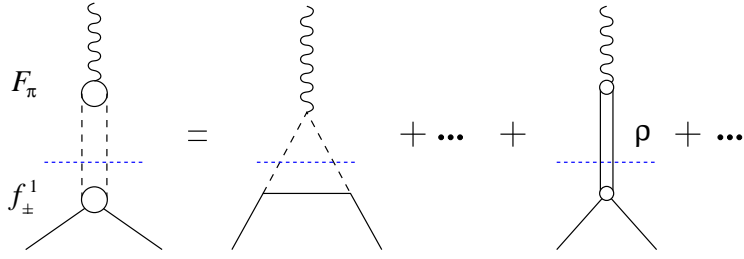}
}
\caption{Two--pion contribution to the isovector nucleon form factors. On the 
left side, the exact representation based on unitarity is shown, whereas the 
triangle diagram on the right side leads to the strong enhancement of the 
isovector spectral functions close to threshold. Also shown is the dominant 
$\rho$--meson contribution. 
The solid, dashed, wiggly and double lines
represent nucleons, pions, photons and the $\rho$, respectively.
\label{fig:2pic}} 
\end{figure}

\begin{figure}[b] 
\centerline{\includegraphics*[width=8cm,angle=0]{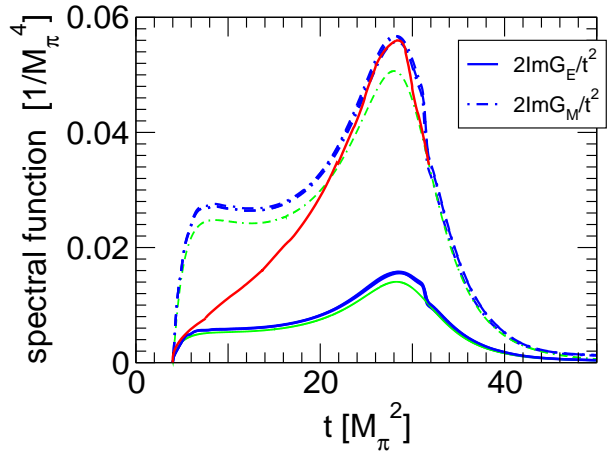}}
\caption{
The two-pion spectral function using the new high statistics data for 
the pion form factor \cite{CMD2,KLOE,SND}. The spectral functions weighted by 
$1/t^2$ are shown for $G_E$ (solid line) and $G_M$ (dash-dotted line). 
The previous results by H\"ohler et al.~\cite{LB} (without $\rho$-$\omega$ 
mixing) are shown for comparison by the gray/green lines. The red solid line
indicates the $\rho$-meson contribution to Im~$G_M$ with a width 
$\Gamma_\rho = 150\,$MeV.}
\label{2pispec}
\end{figure}

\medskip\noindent
{\bf Isovector case:}  Let us now  evaluate the two--pion contribution in a 
model--independent way and draw some conclusions on the spatial extent of the 
pion cloud from that (see the next section). 
As pointed out long ago~\cite{FF} and further elaborated 
on in Ref.~\cite{HP}, unitarity allows us to determine the isovector spectral 
functions from threshold up to masses of about 1~GeV in terms of the pion 
charge form factor $F_\pi (t)$ and the P--wave $\pi\pi \bar N N$ partial 
waves, see Fig.~\ref{fig:2pic}. We use here the form 
\beq 
{\rm Im}~G_E^{V} (t) = \frac{q_t^3}{m\sqrt{t}}\, 
\left|F_\pi (t)\right|^2 \, J_+ (t)~,~~~ 
{\rm Im}~G_M^{V} (t) = \frac{q_t^3}{\sqrt{2t}}\, 
\left|F_\pi (t)\right|^2 \, J_- (t)~, 
\label{uni} 
\eeq
where $q_t=\sqrt{t/4-M_\pi^2}$. The functions $J_{\pm} (t)$ are related to the 
$t$--channel P--wave $\pi$N partial waves $f_\pm^1(t)$ via 
$f_\pm^1(t) = F_\pi (t)\, J_\pm (t)$ in the conventional isospin 
decomposition, with the tabulated values of the $J_\pm (t)$ from \cite{LB}. For 
the pion charge form factor $F_\pi$ we use the latest experimental data 
from CMD-2 \cite{CMD2}, KLOE \cite{KLOE}, and SND \cite{SND}.
We stress that the representation of Eq.~(\ref{uni}) gives the {\em exact isovector 
spectral functions} for $4M_\pi^2 \leq t \leq 16 M_\pi^2$ but in practice holds 
up to $t \simeq 50 M_\pi^2$. It has two distinct features. First, as already 
pointed out in \cite{FF}, it contains the important contribution of the 
$\rho$--meson (see Fig.~\ref{fig:2pic}) with its peak at $t \simeq 30 M_\pi^2$. 
Second, on the left shoulder of the $\rho$, the isovector spectral functions 
display a very pronounced enhancement close to the two--pion threshold,
as shown in Fig.~\ref{2pispec} (taken from Ref.~\cite{BHM2pi}). 
\noindent
This is due to the logarithmic singularity on the second Riemann sheet located at 
$t_c = 4M_\pi^2 - M_\pi^4/m^2 = 3.98 M_\pi^2$, very close to the threshold. 
This pole comes from the projection of the nucleon Born graphs, or in modern 
language, from the triangle diagram also depicted in Fig.~\ref{fig:2pic}
(middle graph). If 
one were to neglect this important unitarity correction, one would severely 
underestimate the nucleon isovector radii \cite{HP2}. 
In fact, precisely the 
same effect is obtained at leading one--loop accuracy  in chiral perturbation 
theory, as  discussed first in \cite{GSS,BKMrev}. This topic was further 
elaborated on in 
the framework of heavy baryon CHPT \cite{BKMspec,Norbert} and in a covariant 
calculation based on infrared regularization \cite{KM} (see also
\cite{Schindler:2005ke}). It is  important to note that there is
a strict one-to-one correspondence between this unitarity correction and 
the field-theoretically defined one-pion loop -- contrary to what 
is claimed in de Jager's contribution to
this workshop \cite{deJager:2006nt}. Stated differently, the most 
important two--pion contribution to the nucleon form factors can be determined 
by using either unitarity or CHPT (in the latter case, of course, the $\rho$ 
contribution is not included). Clearly, this is an important input into the 
spectral functions used in the on-going  dispersive analysis of the nucleon 
form factors by the Bonn-Mainz group \cite{MMD,HMD,HM,BHMff} (see also the
work by Dubnicka and collaborators as reviewed in Ref.~\cite{Adamuscin:2005aq}).

\begin{figure}[t]
\resizebox{0.3\textwidth}{!}{%
  \includegraphics{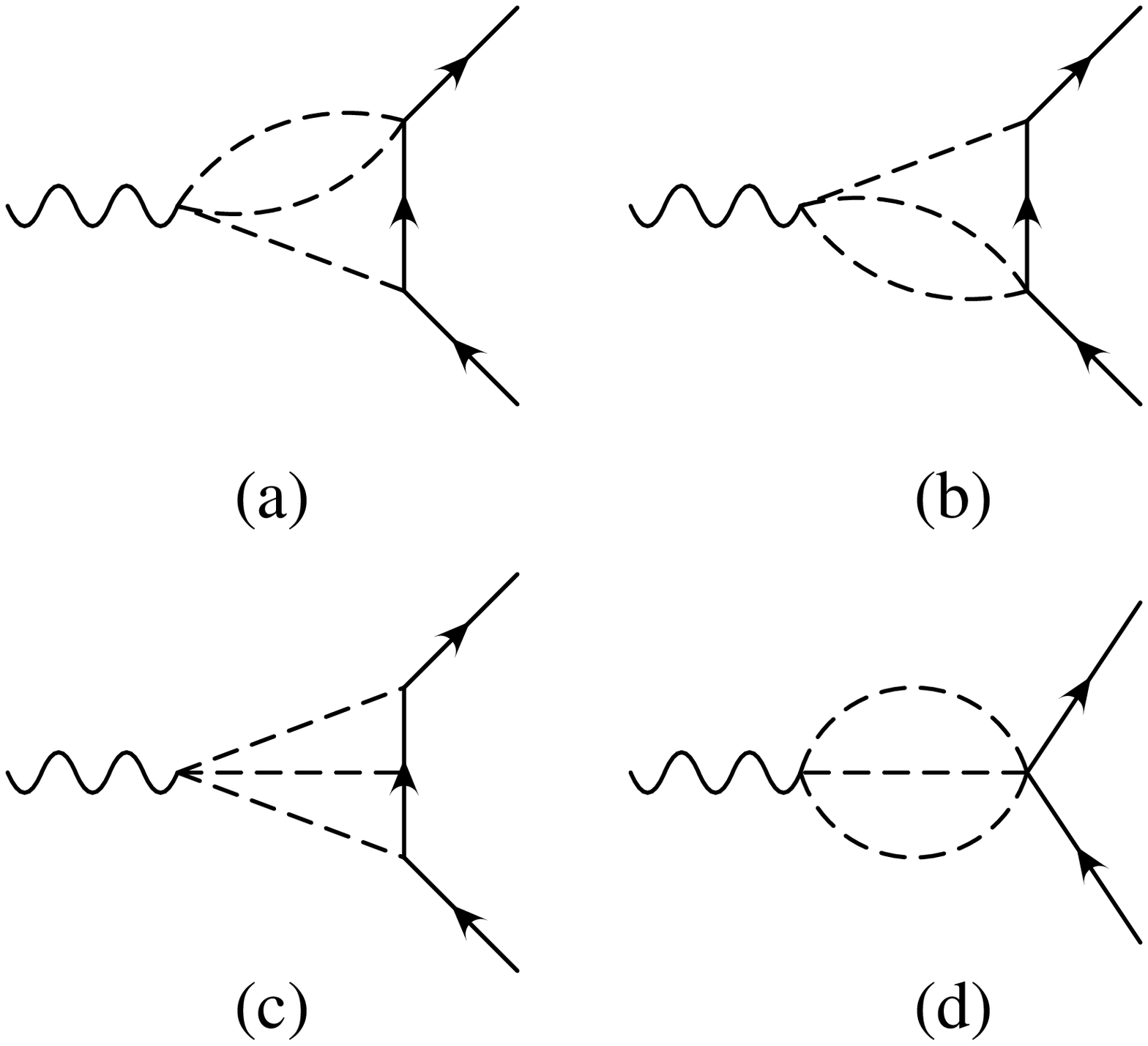}
}
~~~~~~~~~~~~~~~~~~~~\resizebox{0.3\textwidth}{!}{%
  \includegraphics{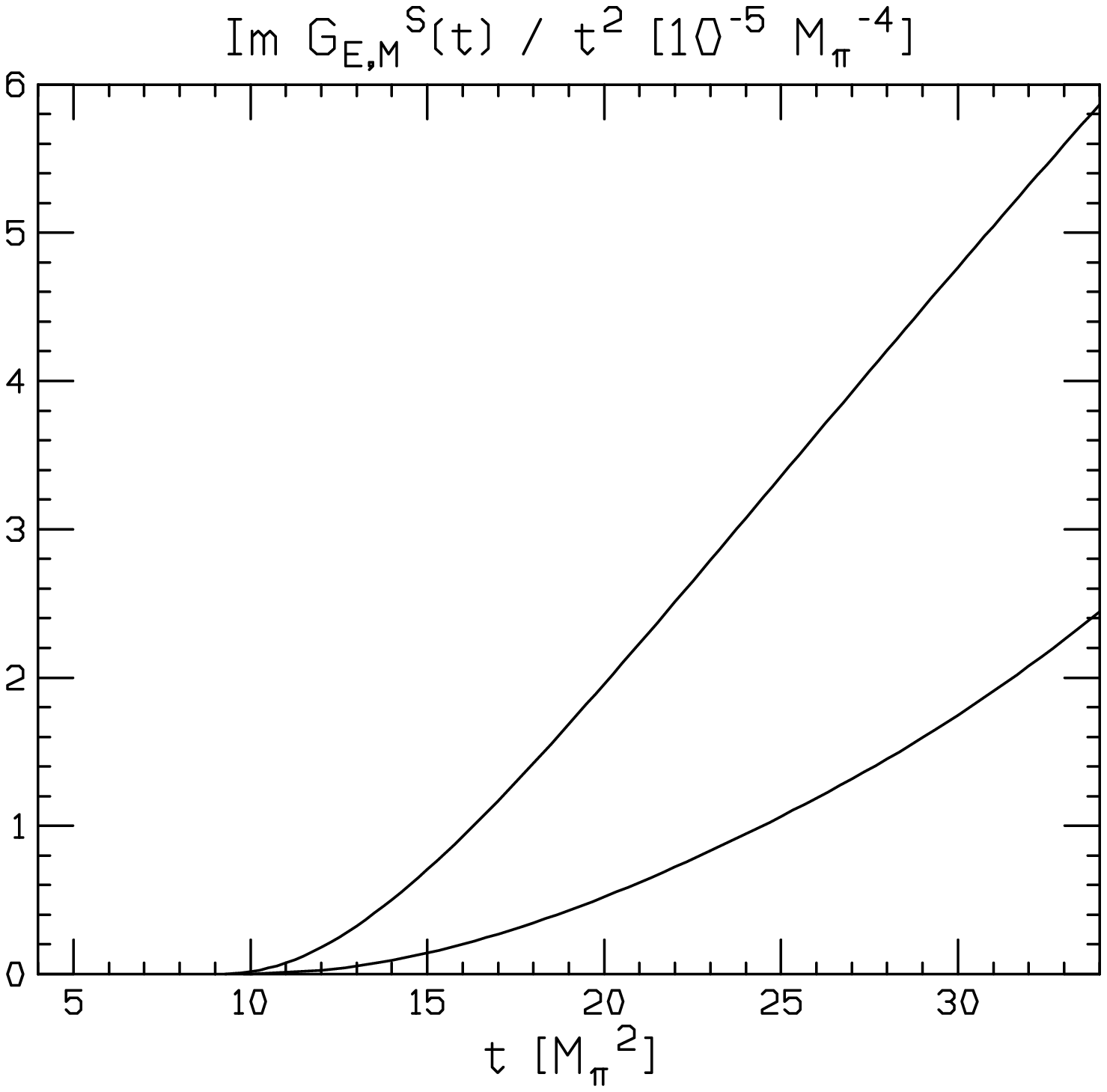}
}
\caption{
Left panel: Two--loop diagrams contributing to the imaginary parts
of the isoscalar electromagnetic nucleon form factors. The solid, dashed and
wiggly lines denote nucleons, pions and photons, respectively. Right panel:
Spectral distribution of the isoscalar electric and magnetic
nucleon form factors weighted with $1/t^2$ in the heavy nucleon limit.
Shown are Im\,$G_M^S (t)/t^2$ (upper line) and Im\,$G_E^S (t)/t^2$ (lower
line).
}
\label{fig:spec3pi}       
\end{figure}

\medskip\noindent
{\bf Isoscalar case:} In the isoscalar electromagnetic channel, it was 
believed (but not proven) that  the pertinent
spectral functions rise smoothly from the three--pion threshold  to the
$\omega$-meson peak, i.e. that there is no pronounced effect from the
three--pion cut on the left wing of the $\omega$-resonance (which also has a
much smaller width than the $\rho$-meson). Chiral perturbation theory 
was used to settle this issue, see Ref.~\cite{BKMspec}. An investigation of the
isoscalar spectral functions based on pion scattering data and dispersion 
theory as done for the isovector  spectral functions 
seems not to be feasible at the moment since it requires the full
dispersion--theoretical analysis of the three-body processes $\pi N  \to \pi\pi
N$ (or of the data on $\bar N N \to 3\pi$). Consider now the CHPT analysis.
The imaginary parts of the isoscalar electromagnetic form factors open at the
three--pion threshold $t_0 = 9\,M_\pi^2$. The leading two--loop diagrams
to the three--pion cut contribution are depicted in the left panel of 
Fig.~\ref{fig:spec3pi}. 
A compact form of the isoscalar spectral functions can be given in the
limit $m\to \infty$. Furthermore, these
results  represent the genuine leading order
contributions with all higher order effects (starting at order $q^8$
in the chiral expansion) switched off,
\begin{equation} 
{\rm Im}\, G_E^S(t) = {3g_A^3 \,t\over (4 \pi)^5 F_\pi^6}
\int\int_{z^2<1} d \omega_1 d\omega_2\, |{\bf{l_1}}| \, |{\bf{l_2}}| \,  
\sqrt{1-z^2} \, \arccos(-z) \,\,,
\label{imges}
\end{equation}
\begin{eqnarray} 
&&{\rm Im}\, G_M^S(t) = {g_Am\over (8 \pi)^4 F_\pi^6}
\biggl\{L(t) \biggl[ 3t^2 -10 tM_\pi^2 + 2M_\pi^4 + g_A^2
\bigl(3t^2 -2 tM_\pi^2 - 2M_\pi^4 \bigr)\biggr]\nonumber\\ && \quad+W(t)\biggl[
t^3+2t^{5/2}M_\pi-39t^2 M_\pi^2 -12 t^{3/2}M_\pi^3 +65 t M_\pi^4 -50 
\sqrt t M_\pi^5 -27 M_\pi^6 \nonumber \\ && \quad +g_A^2 \bigl( 5t^3+10t^
{5/2}M_\pi- 147t^2 M_\pi^2 +36 t^{3/2}M_\pi^3 +277 t M_\pi^4 -58 \sqrt
t M_\pi^5 -135 M_\pi^6 \bigr) \biggr] \biggr\}\,\,,
\label{imgms} 
\end{eqnarray}   
with
\beq 
L(t) = {M_\pi^4 \over 2t^{3/2}} \ln {\sqrt
t-M_\pi+\sqrt{t -2\sqrt t M_\pi-3M^2_\pi}\over2M_\pi}\,\, , ~~
W(t) = {\sqrt t- M_\pi\over 96 t^{3/2}}\sqrt{t-2\sqrt t M_\pi-
3M^2_\pi} \,\,\, , 
\eeq
and ${\bf {l}_{1,2}}$ are the two independent pion momenta of the 
three-particle intermediate state (for precise definitions, see
Ref.~\cite{BKMspec}).
Here, $g_A$ is the nucleon axial-vector coupling and $F_\pi$ the
pion decay constant.
Note that in the infinite nucleon mass limit Im\,$G_E^S(t)$ comes solely from
graph (c) in Fig.~\ref{fig:spec3pi} and quite astonishingly one can evaluate 
all integrals in closed form for Im\,$G_M^S(t)$. 
The behavior near threshold $t_0 = 9\, M_\pi^2$ of the imaginary parts
for finite pion mass is 
\begin{equation} 
{\rm Im}\, G^S_E(t) \sim (\sqrt t-3M_\pi)^3\,, \qquad {\rm
Im}\, G^S_M(t) \sim (\sqrt t-3M_\pi)^{5/2} 
\end{equation}
which corresponds to a stronger growth than pure phase space.
This feature indicates (as in the isovector case) that in the heavy nucleon
mass limit $m \to \infty$ normal and anomalous thresholds coincide. In order to
find these singularities for finite nucleon mass $m$ an investigation of the 
corresponding Landau
equations is necessary \cite{polk}.  By using standard techniques
we are able to find (at least) one anomalous threshold of diagrams (a)
and (b) at  
\begin{equation} 
\sqrt{t_c} = M_\pi \biggl(\sqrt{4-M_\pi^2/m^2}+
\sqrt{1-M_\pi^2/m^2} \biggr)\,\,, \quad \qquad t_c = 8.90\, M_\pi^2~,
\label{anthr}
\end{equation} 
which is very near to the (normal) threshold $t_0=9\, M_\pi^2$ and indeed
coalesces with $t_0$ in the infinite nucleon mass limit. We note that diagram
(d) does not possess this anomalous threshold $t_c=8.90\,M_\pi^2$, but only
the normal one. We do not want to go here deeper into the rather complicated
analysis of the full singularity structure of all two-loop diagrams 
but are mainly interested in the magnitude of the
isoscalar electromagnetic imaginary parts. 
The resulting spectral distributions again weighted with $1/t^2$ are
shown in the right panel of  Fig.~\ref{fig:spec3pi}.
They show a smooth rise and are two orders of magnitude
smaller than the corresponding isovector ones. This
smallness justifies the procedure in the dispersion--theoretical
analysis like in \cite{MMD,HMD,HM,BHMff} to describe the isoscalar spectral
functions solely by vector meson poles starting with the $\omega$-meson 
in the low--energy region. Nevertheless, it may be worthwhile to include these
calculated isoscalar imaginary parts in future dispersion analyses.  
We finally remark that Im~$G_{E,M}^S(t)/t^4$
which have the same asymptotic behavior (for $t\to \infty$) as
Im~$G^V_{E,M}(t)/t^2$ (considering only the leading $q^3$ contribution) do
still not show any strong peak below the $\omega$--resonance.  
Im~$G^S_E(t)/t^4$
is monotonically increasing from $t_0= 9\,M_\pi^2$ to $t=30\,M_\pi^2$ and
Im~$G_M^S(t)/t^4$ develops some plateau between $t= 20$ and $30 \,M_\pi^2$.    
This observation is a further indication  that there is indeed no enhancement 
of the isoscalar electromagnetic spectral function near threshold. 
Even though the isoscalar and isovector
electromagnetic form factors behave formally very similar concerning the 
existence of anomalous thresholds $t_c$ very close to the normal thresholds
$t_0$, the influence of these on the physical spectral functions is
rather different for the two cases. Only in the isovector case a strong
enhancement is visible. This is due to the different phase space
factors, which are $(t-t_0)^{3/2}$ and $(t-t_0)^4$ for the isovector
and isoscalar case, respectively. In latter case,  the
anomalous threshold at $t_c = 8.9\,M_\pi^2$ is thus effectively masked.

\section{The pion cloud as seen in the isovector nucleon form factors}
\label{cloudff}

To get a semi-quantitative idea about the size of the pion cloud
in the nucleon  electromagnetic form factors,
let us separate the (uncorrelated) pion contribution from 
the $\rho$--contribution in the isovector spectral functions
\cite{HMDcloud}.
For  that we decompose the isovector spectral functions as 
\beq 
\label{separation} 
{\rm Im}~G_I^V (t) = {\rm Im}~G_I^{V,2\pi} (t) + {\rm Im}~G_I^{V,\rho} (t)~, 
\quad I = E, M ~, 
\eeq 
and analogously for Im~$F_{1,2}^V (t)$. Using  Eq.~(\ref{dens}), we can then 
calculate the pion cloud contribution to the charge and magnetization density 
in the Breit--frame. The $\rho$--contribution in Eq.~(\ref{separation}) can be 
well represented by a Breit--Wigner form with a running width \cite{Norbert}, 
\beq
\label{rho} 
{\rm Im}~G^{V,\rho}_I (t) = \frac{b_I M_\rho^2 \sqrt{t} 
\Gamma_\rho (t)}{(M_\rho^2-t)^2 + t \Gamma_\rho^2 (t)}~, \quad I = E, M~, 
\eeq 
with the mass $M_\rho = 769.3\,$MeV and the width 
$\Gamma_\rho (t) = g^2(t-4M_\pi^2)^{3/2}/(48 \pi t)$, where the coupling 
$g = 6.03$ is determined from the empirical value 
$\Gamma_\rho (M_\rho^2) = 150.2\,$ MeV, and the parameters $b_I$ can be 
adjusted to the height of the resonance peak. The corresponding expressions 
for the imaginary parts of the Dirac and Pauli form factors can be obtained 
from Eq.~(\ref{conv}). It is clear that the separation into the (uncorrelated) 
pion contribution and the $\rho$--contribution introduces some 
model--dependence. To get an idea about the theoretical error induced by this 
procedure, we perform the separation in three different ways: 
\begin{itemize}
\item[(a)] The two--pion contribution can be directly obtained from the 
two--loop chiral perturbation theory calculation of \cite{Norbert}. Together 
with the $\rho$--contribution of Eq.~(\ref{rho}), this calculation gives a 
very good description of the empirical spectral functions. Note that 
on the right side of the $\rho$, the two--loop representation is slightly 
larger than the empirical one, so that we expect to obtain an upper bound by 
employing  this procedure. We will use the analytical formulae given in 
\cite{Norbert} where the low--energy constant $c_4$ was readjusted to 
avoid double counting of the $\rho$--contribution (see \cite{BKMlecs}). 
\item[(b)] A lower bound on the two--pion  contribution can be obtained by 
setting $F_\pi (t) = 1$ in  Eq.~(\ref{uni}). This prescription does not 
only remove the $\rho$--pole but also some small uncorrelated two--pion 
contributions contained in the pion form factor. 
\item[(c)] To obtain the two--pion contribution, we can also subtract 
Eq.~(\ref{rho}) from the spectral function Eq.~(\ref{uni}) including the full 
pion form factor. The 
parameters $b_E=1.512$ and $b_M=5.114$ are determined such that the two--pion 
contribution at the $\rho$--resonance matches the two--loop chiral 
perturbation theory calculation of \cite{Norbert}. Variation of the $b_I$ 
around these values gives an additional error estimate. 
Note that a similar procedure was performed in \cite{UGMscal} 
to extract scalar meson properties from the scalar pion form factor. 
\end{itemize} 
Using these three methods, we obtain a fairly good handle on the 
theoretical accuracy of the non-resonant two--pion contribution. 
We can now work out the density 
distribution of the two--pion contribution to the nucleon electromagnetic 
form factors. 
Before showing the results, some remarks are in order. As stated 
above, the spectral functions are determined by unitarity (or chiral 
perturbation theory) only up to some maximum value of $t$, denoted
$t_{\rm max}$ in the following. Thus, we have 
simply set the spectral functions in the integral Eq.~(\ref{dens}) to zero for 
momentum transfers beyond the value $t_{\rm max} = 40 M_\pi^2$. 
\begin{figure}[t]
\resizebox{0.45\textwidth}{!}{%
  \includegraphics{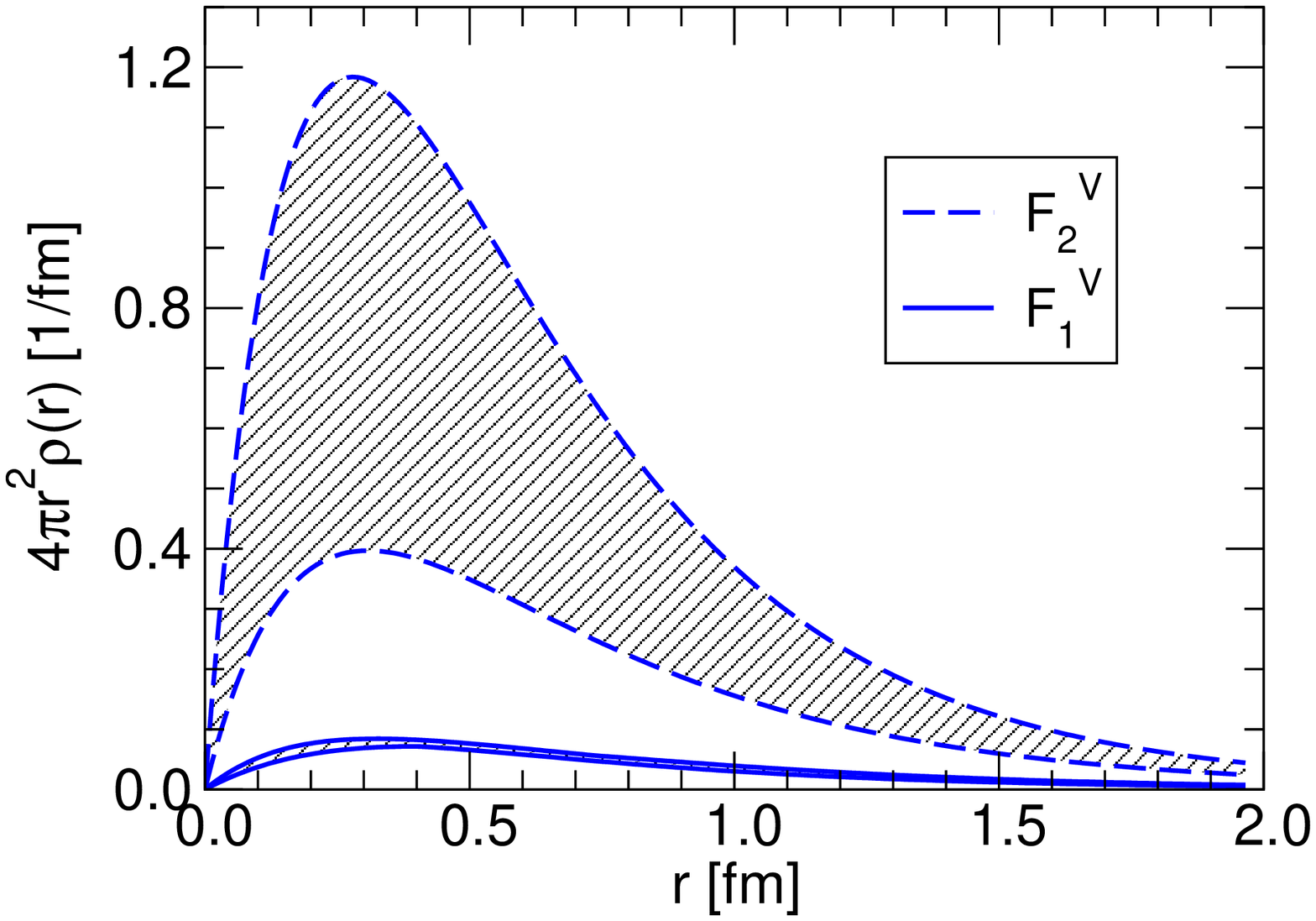}
}
~~~~~~~~~~~~~\resizebox{0.45\textwidth}{!}{%
  \includegraphics{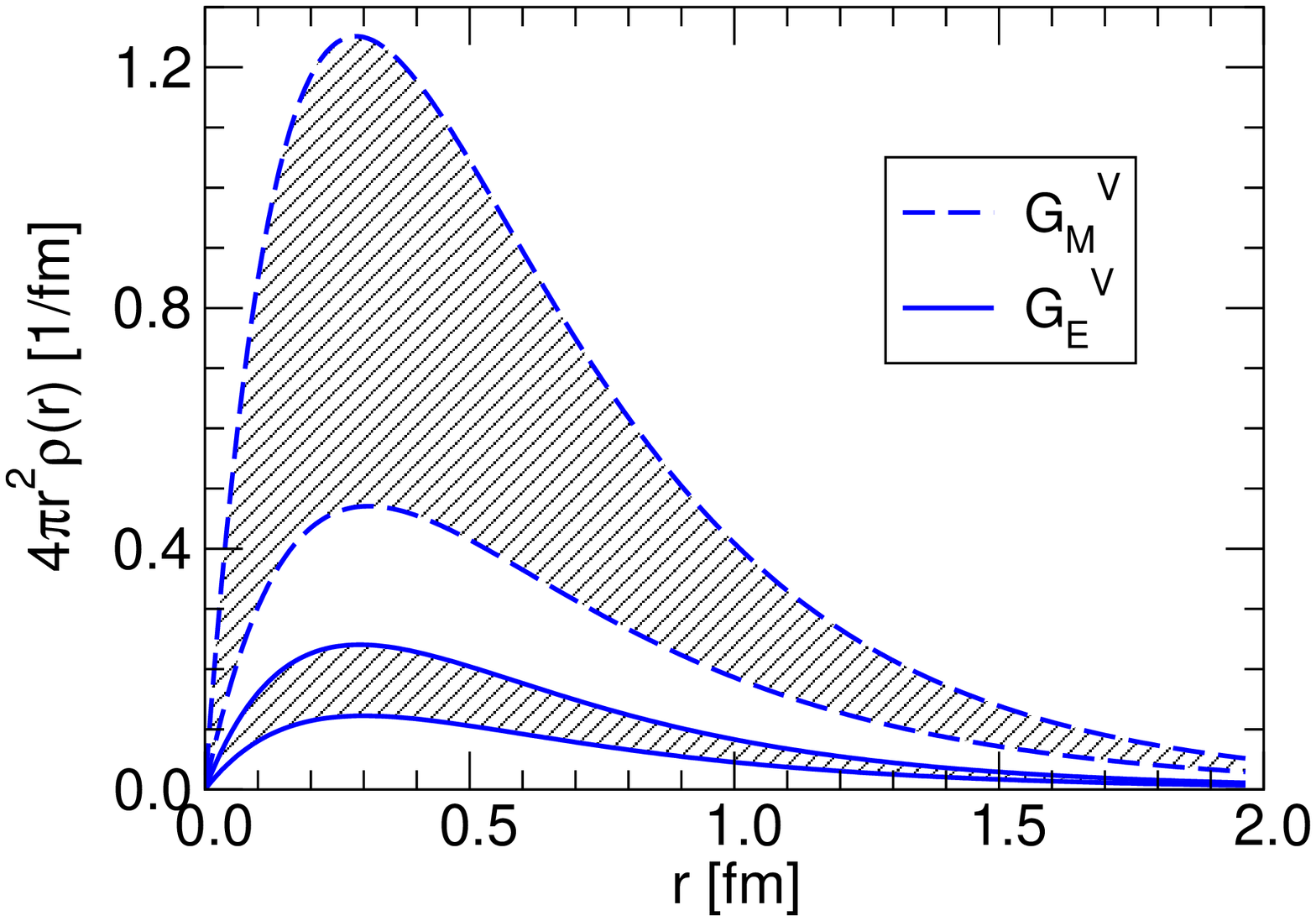}
}
\caption{
The densities of charge and magnetization due to the pion cloud. Left 
panel: $4\pi r^2 \rho(r)$ for the isovector Pauli (upper band) and Dirac 
(lower band) form factors. Right panel: $4\pi r^2 \rho(r)$ for the isovector 
magnetic (upper band) and electric (lower band) Sachs form factors. 
}
\label{fig:FF}       
\end{figure}
In Fig.~\ref{fig:FF}, we show the resulting densities for the isovector form 
factors weighted with $4\pi r^2$. The contribution of the ``pion cloud'' to 
the total charge or magnetic moment is then simply obtained by integration 
over $r$. The bands reflect the theoretical uncertainty in the separation. For 
all form factors, the lower and upper bounds are given by methods (b) and (a), 
respectively. Method (c) generally yields a result between these bounds,
except for the Dirac form factor where it gives the upper bound. The weighted 
densities for the isovector Dirac and Pauli form factors are shown in the left 
panel of Fig.~\ref{fig:FF}. We see that these charge distributions show a 
pronounced peak around $r \simeq 0.3\,$fm, quite consistent with earlier 
determinations (see e.g. \cite{UGM,Holz}), and fall off smoothly with 
increasing distance. In the right panel of Fig.~\ref{fig:FF}, we show the 
densities (again weighted with $4\pi r^2$) for the electric and magnetic Sachs 
form factors which come out very similar to the case of the Dirac and  Pauli 
form factors. In comparison with Ref.~\cite{FW}, we generally obtain much 
smaller pion cloud effects at distances beyond 1~fm, e.g., 
by a factor 3 for $\rho_E^V (r)$ at  $r = 1.5$~fm.
We have also studied the sensitivity of our results to the cut--off 
$t_{\rm{max}}$. While this may increase the value of the ``pion cloud'' 
contribution, it leaves the position of the maximum essentially unchanged. 
However, it is obvious from Eq.~(\ref{dens}) that masses beyond 0.5~GeV  and 
corresponding small--distance phenomena ($r \leq 0.4$~fm) should not be 
related to the pion cloud of the nucleon. Finally, we show the corresponding 
two--pion contribution to the charges and radii for the various nucleon form 
factors in Table~\ref{radtab}. 
The contribution of the pion cloud to the isovector electric (magnetic) 
charge is 20\%~(10\%)
in the model of Ref.~\cite{FW}. This is consistent with our range of values 
for the electric charge but a factor of 1.5 smaller than our lower bound
for the magnetic one, see Table~\ref{radtab}.
Furthermore, note that the pion cloud gives only a fraction of all form factors
at zero momentum transfer. Normalized to the contribution of the 
pion cloud, the corresponding radii are of the order of 1~fm. In the model
of \cite{FW}, these radii are considerably larger, of the order of 1.5~fm.
Note that if one shifts all the strength of the corresponding spectral 
functions to threshold, one obtains an upper limit $r_{\rm max } = \sqrt{3/2}\,
M_\pi^{-1} \simeq 1.7\,$fm, assuming that the spectral functions are positive 
definite.
%
\begin{table}[t!]
\caption{Two--pion contribution to charges and radii (in fm$^2$)
for the various nucleon form factors. The radii are normalized to the 
physical charges and magnetic moments.} 
\label{radtab}       
\begin{tabular}{cccccccc} 
\hline
$F_1^V(0)$ & $F_2^V(0)$ & $G_E^V(0)$ & $G_M^V(0)$ & 
$\langle r^2\rangle_1^V$ & $\langle r^2\rangle_2^V$  & 
$\langle r^2\rangle_E^V$ & $\langle r^2\rangle_M^V$ \\ 
\hline
$0.07...0.08$ & $0.4...1.0$ & $0.1...0.2$ & $0.4...1.0$ & $0.1...0.2$ 
& $0.2...0.3$ & $0.2...0.3$ & $0.2...0.3$ \\ 
\hline
\end{tabular}
\end{table}

\section{What can we conclude?}
\label{conclusions}
Let me summarize the pertinent conclusions of this talk:
\begin{itemize}
\item[i)] Chiral perturbation theory is the natural framework to
investigate the role of pionic contributions to hadron (nucleon)
structure. Nucleon observables receive contributions from pion loops,
the ``pion cloud``.
\item[ii)] In general at any given order  in the chiral expansion beyond
tree level, S-matrix elements and transition currents receive contributions 
from pion loops and local short-distance operators. Both these contributions
are in general scale-dependent and thus it is possible to shuffle strength from one
to the other. Consequently, an unambiguous extraction of the pion cloud
contribution is not possible.
\item[iii)] The spectral functions that parameterize the physics of the
isovector and isoscalar nucleon electromagnetic form factors are dominated 
for low masses by two- and three-pion exchanges, respectively. The two-pion
contribution can be exactly worked out up to approximately 1~GeV by unitarity
in terms of $\pi N$ scattering amplitudes and the pion vector form factor.
The CHPT representation shares the same analytic properties, namely the
strong enhancement on the left shoulder of the $\rho$ due to an anomalous
threshold on the second sheet close to the physical threshold. A similar
anomalous threshold effect in the isoscalar spectral functions is washed out
by phase space factors. These constraints must be included in any serious 
analysis of the nucleon form factors.
\item[iv)] A model-dependent separation of the correlated from the
uncorrelated two-pion exchange allows one to analyze the spatial extent of this
longest range contribution to the isovector form factors. It is much more
confined in space than in the analysis of Ref.~\cite{FW}.
\item[v)] Concerning low momentum transfer bump-dip structures in the nucleon
form factors for low $Q^2$, one should first realize that such structures
have already been present in most dispersive analyses in the magnetic form
factors. The novel structure in $G_E^n (Q^2)$ proposed in Ref.~\cite{FW}
can only be explained with spectral functions that contain additional 
light mass poles violating the strictures from unitarity and chiral symmetry
as discussed above. According to the newest dispersive analysis \cite{BHMff},
this bump-dip structure lies completely within the one sigma uncertainty and
it requires an additional isoscalar/isovector pole close to the
$\omega$/three-pion threshold. For a more detailed discussion on this topic,
I refer to Ref.~\cite{Hammer:2006mw}.

\end{itemize}

\section*{Acknowledgments}
I am grateful to my collaborators Maxim Belushkin, V\'eronique Bernard, Dieter
Drechsel, Hans-Werner Hammer, Norbert Kaiser and Bastian Kubis. I thank the
organizers for the invitation and excellent organization.



\begin{thebibliography}{1}

\bibitem{Yuk} H.~Yukawa, Proc. Math.-Phys. Soc. Japan {\bf 17}, 48 (1935).

\bibitem{FHK}H. Fr\"ohlich, W. Heitler and N. Kemmer,
 Proc. Roy. Soc. A {\bf 166 }, 155 (1938).\vs

\bibitem{BMrev}
V.~Bernard and U.-G.~Mei{\ss}ner,
   Ann.\ Rev.\ Nucl.\ Part.\ Sci.\
   {\bf 57} (2007) in print [hep-ph/0611231].

\bibitem{Georgi:1994qn}
H.~Georgi,
Ann.\ Rev.\ Nucl.\ Part.\ Sci.\  {\bf 43}, 209 (1993).

\bibitem{Espriu:1993if}
  D.~Espriu and J.~Matias,
  Nucl.\ Phys.\ B {\bf 418}, 494 (1994)
  [arXiv:hep-th/9307086].

\bibitem{Ecker:1988te}
  G.~Ecker, J.~Gasser, A.~Pich and E.~de Rafael,
  Nucl.\ Phys.\ B {\bf 321}, 311 (1989).

\bibitem{Ecker:1989yg}
  G.~Ecker, J.~Gasser, H.~Leutwyler, A.~Pich and E.~de Rafael,
  Phys.\ Lett.\ B {\bf 223}, 425 (1989).

\bibitem{Donoghue:1988ed}
  J.~F.~Donoghue, C.~Ramirez and G.~Valencia,
  Phys.\ Rev.\ D {\bf 39}, 1947 (1989).

\bibitem{Kampf:2006bn}
  K.~Kampf and B.~Moussallam,
   Eur.\ Phys.\ J.\ C {\bf 47}, 723 (2006)
  [arXiv:hep-ph/0604125].


\bibitem{Bernard:2003rp}
  V.~Bernard, T.~R.~Hemmert and U.-G.~Mei{\ss}ner,
  Nucl.\ Phys.\ A {\bf 732}, 149 (2004)
  [arXiv:hep-ph/0307115].


\bibitem{MMD} P.~Mergell, U.-G.~Mei{\ss}ner and D.~Drechsel, 
  Nucl.\ Phys.\ A {\bf 596}, 367 (1996) [arXiv:hep-ph/9506375].\vs 

\bibitem{FF} W.~R.~Frazer and J.~R.~Fulco, 
  Phys.\ Rev.\ Lett.\ {\bf 2}, 365 (1959).\vs 

\bibitem{HP} G.~H\"ohler and E.~Pietarinen, 
  Nucl.\ Phys.\ B {\bf 95}, 210 (1975).\vs 

\bibitem{LB} G. H\"ohler, 
  "Pion-Nucleon Scattering", Landolt-B\"ornstein Vol. I/9b, ed. H. Schopper, 
   Springer, Berlin, 1983.\vs 

\bibitem{CMD2}
R.~R.~Akhmetshin {\it et al.}  [CMD-2 Collaboration],
Phys.\ Lett.\ B {\bf 527}, 161 (2002) 
[arXiv:hep-ex/0112031];
%
Phys.\ Lett.\ B {\bf 578}, 285 (2004) 
[arXiv:hep-ex/0308008].

\bibitem{KLOE}
A.~Aloisio {\it et al.}  [KLOE Collaboration],
Phys.\ Lett.\ B {\bf 606}, 12 (2005) 
[arXiv:hep-ex/0407048].

\bibitem{SND}
M.~N.~Achasov {\it et al.},
  J.\ Exp.\ Theor.\ Phys.\  {\bf 101}, 1053 (2005)
  [Zh.\ Eksp.\ Teor.\ Fiz.\  {\bf 101}, 1201 (2005)]
  [arXiv:hep-ex/0506076].

\bibitem{BHM2pi}
  M.~A.~Belushkin, H.~W.~Hammer and U.-G.~Mei{\ss}ner,
  Phys.\ Lett.\ B {\bf 633}, 507 (2006)
  [arXiv:hep-ph/0510382].

\bibitem{HP2} G.~H\"ohler and E.~Pietarinen, 
  Phys.\ Lett.\ B {\bf 53}, 471 (1975).\vs 

\bibitem{GSS} J.~Gasser, M.~E.~Sainio and A.~Svarc,
  Nucl.\ Phys.\ B {\bf 307}, 779 (1988).\vs 

\bibitem{BKMrev} V.~Bernard, N.~Kaiser and U.-G.~Mei{\ss}ner, 
   Int.\ J.\ Mod.\ Phys.\ E {\bf 4}, 193 (1995) [arXiv:hep-ph/9501384].\vs 

\bibitem{BKMspec} V.~Bernard, N.~Kaiser and U.-G.~Mei{\ss}ner, 
  Nucl.\ Phys.\ A {\bf 611}, 429 (1996) [arXiv:hep-ph/9607428].\vs 

\bibitem{Norbert} N.~Kaiser, 
  Phys.\ Rev.\ C {\bf 68}, 025202 (2003) [arXiv:nucl-th/0302072].\vs 

\bibitem{KM} B.~Kubis and U.-G.~Mei{\ss}ner, 
  Nucl.\ Phys.\ A {\bf 679}, 698 (2001) [arXiv:hep-ph/0007056].\vs 

\bibitem{Schindler:2005ke}
  M.~R.~Schindler, J.~Gegelia and S.~Scherer,
  Eur.\ Phys.\ J.\ A {\bf 26}, 1 (2005)
  [arXiv:nucl-th/0509005].

\bibitem{deJager:2006nt}
  K.~de Jager,
  arXiv:nucl-ex/0612026.

\bibitem{HMD} H.~W.~Hammer,  U.-G.~Mei{\ss}ner and D.~Drechsel,
  Phys.\ Lett.\ B {\bf 385}, 343 (1996) [arXiv:hep-ph/9604294].\vs 

\bibitem{HM}
  H.~W.~Hammer and U.-G.~Mei{\ss}ner,
  Eur.\ Phys.\ J.\ A {\bf 20}, 469 (2004)
  [arXiv:hep-ph/0312081].

\bibitem{BHMff}
  M.~A.~Belushkin, H.~W.~Hammer and U.-G.~Mei{\ss}ner,
  Phys.\ Rev.\ C (2007) in print; 
  arXiv:hep-ph/0608337.

\bibitem{Adamuscin:2005aq}
  C.~Adamuscin, S.~Dubnicka, A.~Z.~Dubnickova and P.~Weisenpacher,
  Prog.\ Part.\ Nucl.\ Phys.\  {\bf 55}, 228 (2005)
  [arXiv:hep-ph/0510316].

\bibitem{polk} R.J. Eden, P.V. Landshoff, D.I. Olive and
  J.C. Polkinghorne,  {\it The
Analytic S--Matrix} (Cambridge University Press, Cambridge, 1966).

\bibitem{HMDcloud}
  H.~W.~Hammer, D.~Drechsel and U.-G.~Mei{\ss}ner,
  Phys.\ Lett.\ B {\bf 586}, 291 (2004)
  [arXiv:hep-ph/0310240].

\bibitem{BKMlecs} 
  V.~Bernard, N.~Kaiser and U.-G.~Mei{\ss}ner 
  Nucl.\ Phys.\ A {\bf 615}, 483 (1997) [arXiv:hep-ph/9611253].\vs 

\bibitem{UGMscal} U.-G.~Mei{\ss}ner, 
  Comm.\ Nucl.\ Part.\ Phys.\  {\bf 20}, 119 (1991).\vs 

\bibitem{UGM} U.-G.~Mei{\ss}ner, 
  Phys.\ Rept.\ {\bf 161}, 213 (1988).\vs 

\bibitem{Holz} G.~Holzwarth, 
  Z.\ Phys.\ A {\bf 356}, 339 (1996) [arXiv:hep-ph/9606336].\vs 

\bibitem{FW} 
  J.~Friedrich and T.~Walcher, 
  Eur.\ Phys.\ J.\ A {\bf 17}, 607 (2003) [arXiv:hep-ph/0303054].\vs 

\bibitem{Hammer:2006mw}
  H.~W.~Hammer,
  Eur.\ Phys.\ J.\ A {\bf 28}, 49 (2006)
  [arXiv:hep-ph/0602121].


\end{thebibliography}
\end{document}